# Detecting the Trend in Musical Taste over the Decade – A Novel Feature Extraction Algorithm to Classify Musical Content with Simple Features

Anish Acharya[1]
[1]Dept. of Electrical Engineering and Computer Science
Henry Samueli School of Engineering
University of California-Irvine

## Abstract:

This work proposes a novel feature selection algorithm to classify Songs into different groups. Classification of musical content is often a non-trivial job and still relatively less explored area. The main idea conveyed in this article is to come up with a new feature selection scheme that does the classification job elegantly and with high accuracy but with simpler but wisely chosen small number of features thus being less prone to over-fitting. This uses a very basic general idea about the structure of the audio signal which is generally in the shape of a trapezium. So, using this general idea of the Musical Community we propose three frames to be considered and analyzed for feature extraction for each of the audio signal- opening, stanzas and closing and it has been established with the help of a lot of experiments that this scheme leads to much efficient classification with less complex features in a low dimensional feature space thus is also a computationally less expensive method. Step by step analysis of feature extraction, feature ranking, dimensionality reduction using PCA has been carried in this article. Sequential Forward selection (SFS) algorithm is used to explore the most significant features both with the raw Fisher Discriminant Ratio (FDR) and also with the significant eigen-values after PCA. Also during classification extensive validation and cross validation has been done in a monte-carlo manner to ensure validity of the claims.

## 1. Introduction:

For the past few decades audio processing and classification of songs has emerged as an important and interesting field of study. With increasing volume of data being created in the music industry it is often very useful to develop powerful and efficient classification algorithms to classify Music files into different classes of interest. In most of the real world problems we face, we hardly get a precise symbolic representation of a song and one has to deal with only the audio samples obtained from the sampling of the exact waveform. However, these audio samples can’t be used directly in AI application algorithms because of the low level and low density of information contained within them [1].

The first step to deal with this in an AI problem would be to extract meaningful features from the audio samples to get adequate information and employ further processing. After the features are extracted the idea is to select the most efficient and meaningful feature rather than to take all the features in account keeping in mind the case of over-fitting. Thus the idea is to extract meaningful features from the audio signals, then to rank them according to some pre-defined criterion and select the independent and most significant features [2].

All the music Classification tasks performed previously uses complex features like timbre features such as temporal features, energy features, spectral shape features, perceptual features; melody/ harmony features such as pitch function, unfolded or folded function; rhythm features such as tempo, musical patterns etc. And, of course these complex features are necessary as it is extremely difficult to analyze musical content and classify them into different classes using simple statistical features like moments, entropy, spectral density etc. However, the major contribution of this article is to claim that such simple features can also be used to give good classification accuracy if chosen wisely. Thus this paper proposes the idea of segmenting the audio file into three most significant frames namely Opening, Stanzas and Closing and then extracting simple statistical features like different moments, Power Spectral Density, Fano-Factor etc from each of these frames. Extensive experiments are presented to claim the validity of this claim.

One thumb rule of any Pattern recognition problem is to not only choose the features but to choose the most effective one and keeping the number of features to minimal number. There is more than one reason to reduce the number of feature to a sufficient minimum. Features with high correlation coefficient between them, if treated separately increases the complexity without much gain in information. A large number of features are directly translated to a large number of classifier parameters thus might lead to over-fitting. Thus for a finite and usually limited number of training patterns keeping the number of features as small as possible is in line with our goal of designing classifiers with good generalization capabilities [3]. Thus after extracting the features the most important question is to reduce the dimension of feature space in such that minimal information is lost.
Thus as a first step to do so Statistical hypothesis testing and deviation bounds are invoked to eliminate less informative features i.e. easily recognizable ‘bad’ choices at the very beginning.
Now, eliminating features by looking at them is far from optimal and hence as the next step in Feature Selection more elaborate techniques of feature ranking looking at their interdependence are investigated. In this paper Fisher Discriminant Ratio (FDR) has been used as the ranking criterion of the features [4], [5]. Also, Principal Component Analysis (PCA) has also been investigated to select features leading to largest principal components thus leading to minimal information loss in the process of dimensionality reduction. Here in this work a thorough investigation of feature selection has been taken care of to ensure selection of best features as well as removal of redundant or dependent features. The classical method of Sequential Forward Selection (SFS) scheme has been used for that purpose. The same idea of Sequential Forward Selection (SFS) algorithm has been used in case of PCA based feature selection and all the results have been reported.

Indian Music has always been considered as one of the richest and most complex to analyze with so many delicate and complex variants ranging from Hard Rock, Contemporary Rock, Pure Classical, Semi Classical and so many more variants and thus it is very difficult to draw a line between two as the genres are not well defined and has a lot of overlap. Thus even a binary classification of Indian Music is a non-trivial task and that too on the basis of simple statistical features like moments gives very low accuracy. However, it has been shown here that with the help of the proposed Frame Selection and Feature extraction procedure claimed highly accurate results are possible even in low dimensional Feature Space using simple classical statistical Features like moments, Fano Factor, Spectral Density etc. On the process of exploring Indian Music two major observations are also drawn. Firstly, the statistical difference of so called Indian Rock music with Indian Classical music and Secondly, the evolution of Indian Hit Music over the last decade i.e. it has been shown that Indian hit music of the 90's era is statistically differentiable with the nature of hit music of 2000 era. Thus this can be extended to conclude that the musical taste of the Indian Music audience has also changed over the last decade.

The classification of Indian Music between two different decades which leads to an intuition of the musical taste change of the Indian Music audience with good accuracy is as difficult as it sounds. The mere fact that one would expect high overlap between the music of two decades and thus the songs might not be easily separable into two distinct classes and thus it is a non-trivial problem. However, with the proposed scheme seems to achieve good results even in case of such difficult problem.

Finding the best Classifier for the proposed scheme is also very crucial. Thus, extensive experiment has been done with different Classifiers and their corresponding accuracies have been reported. As a first exploration of the proposed scheme the analysis has been only restricted to Supervised Learning. However, exploration of Clustering Algorithm performances and required modifications of the scheme will be very interesting and left as a future work.

The main Classifiers used here are- Linear Discriminant Analysis (LDA), Quadratic Discriminant Analysis (QDA), Naïve Bayes, LDA with Moholanbish distance, k- Nearest Neighborhood (KNN) with different criterions to select the k nearest neighbors viz. Euclidian distance, Cityblock (sum of absolute difference), Cosine similarity, correlation based; Support Vector Machine (SVM) with linear kernel, quadratic kernel, Rbf kernel, polynomial kernel each with three different methods to calculate the separating hyper-plane namely- Quadratic programming (QP), Sequential Minimal Optimization (SMO) and Least Squares (LS). Each of the classification tasks have been done using 500 fold cross validation in Monte-Carlo manner to ensure good generalization of the results.

The rest of the paper is organized as follows: In Section 2 we propose our two claims- Firstly, we talk about the chosen dataset which itself is unique and was poorly understood before. Also, we propose our algorithm of partitioning each song along with all intricate details which is our main claim. In section 3 we talk about feature extraction procedures along with the pre-processing steps and giving an overview of the chosen features and the rationale behind choosing them. Section 4 gives a brief overview on Feature selection schemes in the lights of Hypothesis testing deviation bounds, along with the implementation detail and results. In section 5 we talk about different feature ranking algorithms, especially those used in this paper and also provide implementation detail of our chosen algorithm. After gaining detailed knowledge about the Statistics of music we shift to Machine Learning regime in the next section. This section talks about different commonly used supervised learning techniques and also talks about the chosen classifiers along with implementation detail and extensive experimental results. Section 7 concludes and summarizes our findings in the paper and also gives some intuition about the future line of work on the problem. The Reference section provides a thorough citation of all the relevant work.

## 2. Introduction to the Dataset and proposing a novel Feature Extraction scheme:

In the regime of Music processing it is quite common to extract statistical features that are too complicated and not straight forward statistical measures. They also rely on tags like artist name, song title etc. For example Million Song Dataset [9] a very commonly used quite popular dataset in the Music Information Retrieval Community provides many features like song title, artist name, artist term frequency, track ID etc which might be useful in building recommender systems based on collaborative filtering however does not tell much about the statistics of the music and we believe that Music does not have any geographical or language boundary and can be classified only through statistical analysis of the music signal. The statistical features used here are not straight forward and mostly complex or hybrid features for example timber, pitch, loudness, tempo etc which are fine in case of musical sense however they don't generalize to basic statistical features. They also use some loosely defined features like artist hotness, dance-ability, beat confidence, song hotness etc which according to us cannot be a good indicator of musical content. One might conclude and it is quite natural that it is tremendously difficult to analyze musical content only in terms of standard statistical features like standard moments, spectral density etc is very difficult and these statistical features are not often significant and do not have good discrimination power.

Thus now we can understand from the outset that classifying music based on simple statistical features only is not simple and often prove to be insignificant. However, this article proposes a new feature extraction scheme which has been proved to increase the discrimination power of these simple statistical features and thus this paper claims that if properly chosen then even these simple statistical features can lead to excellent classification results as have been proven with extensive experiments in subsequent parts of this paper.

The main idea is based on a simple fact, well known but often ignored in Music Information Retrieval and have never been exploited before in any literature regarding Musical Information Retrieval before. It is well known in the Musical Community that the time domain waveform of any song falls into the envelope of a trapezium. Thus, if we look at the time domain amplitude envelope of a song it should look like a trapezium called Trapezoidal envelope and has been reported in various literatures [6][7] and the waveform looks like as shown in Fig.1.and has three main distinguishable part namely – Opening/ Attack, Stanzas/ Sustain, Closing/Decay. The same notion is known as Mukhra, Alap and Antara in Indian Music. So, the proposal is to select three zones of a music signal according to this notion and treat each Frame as individual entities. The main intuition behind this is that instead of analyzing the entire music signal as a whole it is more reasonable to look into these three frames as it is more likely that these three frames will provide better intuition about the statistics of the song. Thus, instead of extraction statistical features from the entire song we rather extract features from each of these frames and treat them as new features so though our feature space increases three folds but still intuitively we are more likely to extract more information and even simple features like moments of a particular frame might prove to be and is expected to capture more information and turn out to be of higher discriminating power.

Indian Music is considered as one of the richest and versatile in the world with so many variants which range from purely Classical to Rock however if we look into the current genre classification of Indian music there is no existing genre classification which can successfully classify Indian music in different categories. The main reason behind that might be the fact that Indian Music tends to be too complex to be classified and as a result it's like fusion of commonly claimed genres like Rock, pop, jazz, hip-hop etc statistically. Though they sound entirely different but if we look deeply into the statistics of Indian Music a proper classification even into two separate classes is a non-trivial task as everything seems overlapping with the distinction in mannerism of the song which can be very difficult to understand in terms of simple statistical features. Thus, as we claim our algorithm is strong enough to classify Complex and highly overlapping Music like Indian Music and also take the opportunity to be the first to perform such tasks on Indian music to categorize into two classes- Old Indian music (1985-1999) and Contemporary Indian Music (2000-2014) and we have chosen around 350 Samples from both the category to analyze. All these songs used in this paper are chosen blindly from the Popular Ratings available in various Music Ranking Websites purely based on popularity. This serves two purposes- Firstly, as can be understood that it is very difficult to classify any kind of music based on era's and thus these broad classes are expected to have significant overlap and may not be easily separable and hence it would serve as a good platform to test our claim of increasing separability of features if chosen accordingly. Secondly, this would be the first attempt to classify the Hit Music of two decades and from that we might be able to conclude on how the taste of the audience have changed over the past decade. Thus this can also be extended to infer about the trend change in Musical Taste over time.

Thus all said and done the first step is to verify whether the assumption of music files having a trapezoidal envelope was true. For that one song from each category is randomly chosen and the time domain amplitude plots are verified to be supportive of the rationale behind the proposal as can be seen from Fig.2 and Fig.3 and can be readily verified to indeed resemble with the trapezoidal envelope shown in Fig.1.

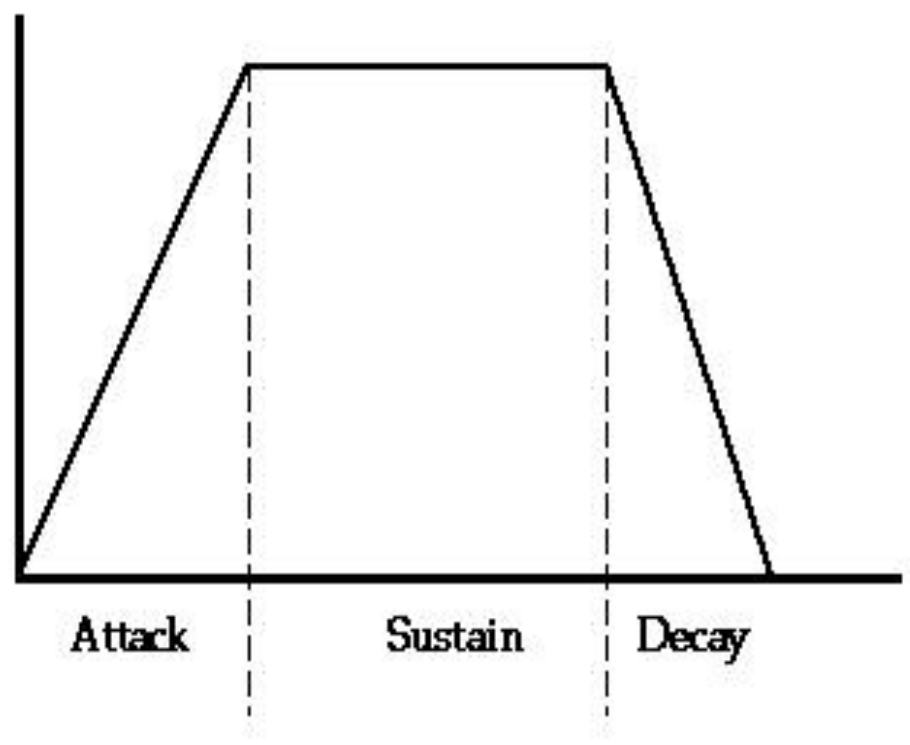

Fig.1. Trapezoidal Envelope

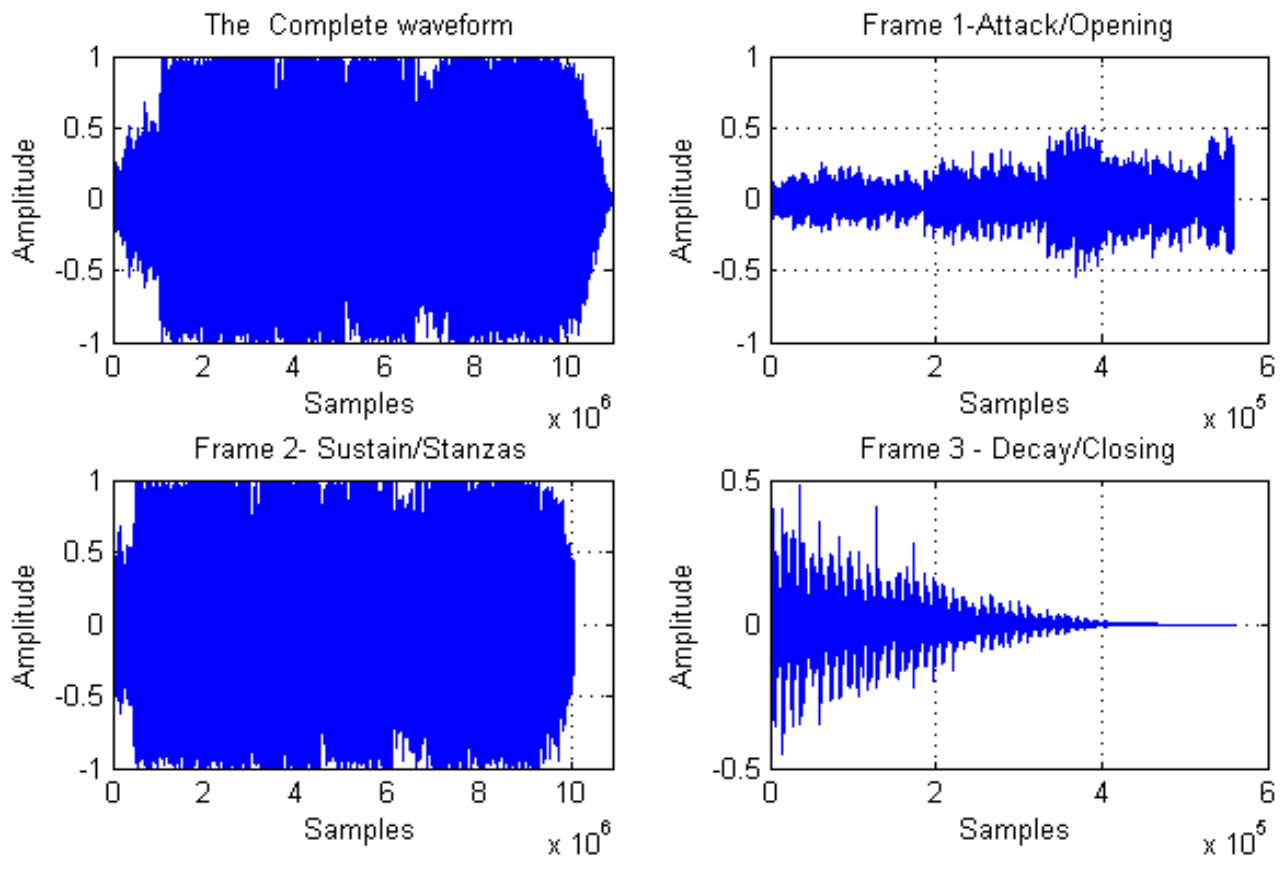

Fig.2. Music of 90s resembles Trapezoidal envelope

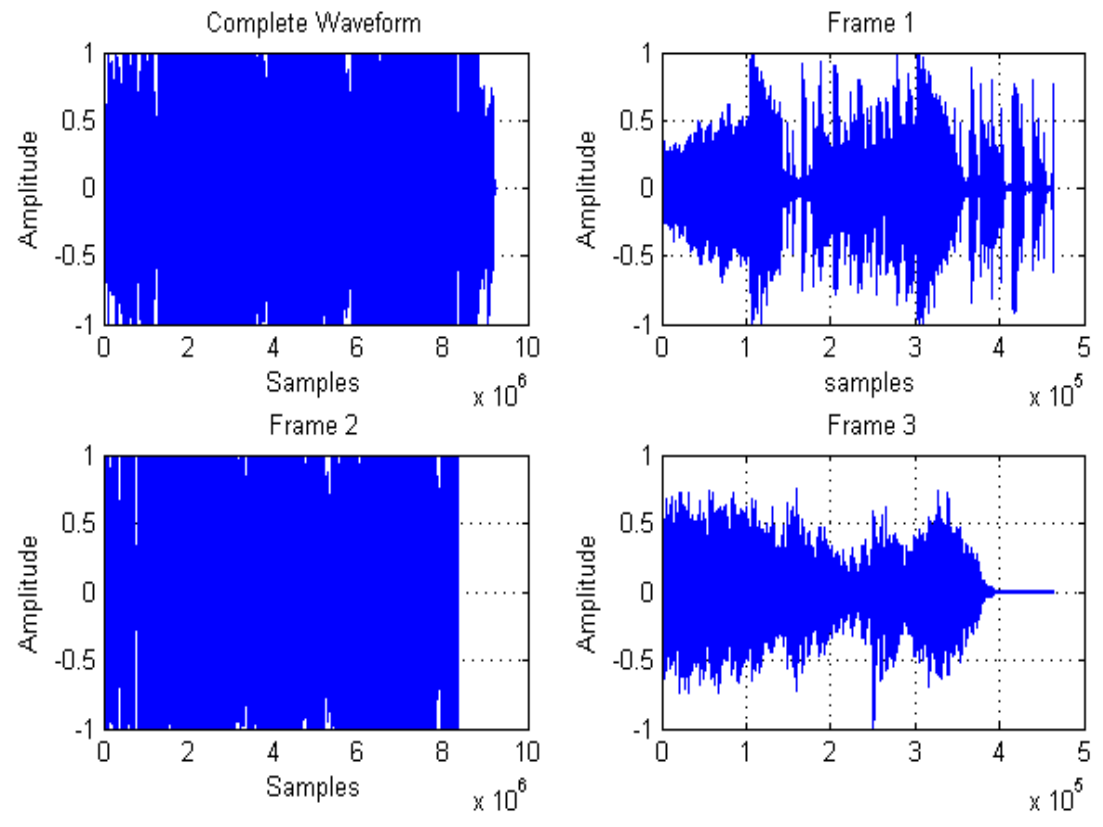

Fig.3. Contemporary Music audio envelope

Thus looking at the two signal envelopes of the time signal of both type of music it can be concluded that both resemble the general pattern of the trapezoidal envelope as shown in fig. 1. However, a close look reveals that each frame of a song belonging to a particular class significantly differs from the corresponding frame of the audio belonging to the other class. Thus as it has been claimed in this paper that dividing each audio file into these three frames and extracting features from them separately looks to be reasonable.

Thus in order to proceed with this new promising idea to divide into these three frames it is very important to comment on the size of each frame. It is very clear that intuitively it is very difficult to propose an accurate exact length of each frame as the duration of each frame might vary from one song to another. Thus to pin point a particular boundary between one frame to another for all songs is not possible. However, it can be seen that intuitively the first 4-6% belong to Frame 1 and last 4-6% approximately belong to Frame 3. Thus, after a few experiments it seems to work best if the First 5% of the samples are considered to belong to Frame 1 i.e. opening and the last 5% belonging to closing or Frame 3. Thus, for the rest of the paper the frame sizes are unanimously distributed as- Frame1 is the first 5% samples of the audio signal, Frame 2 is the next 90% of the audio signal and last 5% of the samples belong to Frame3 as the generalized division rule for all the audio files.

## 3. Feature Extraction and Dataset Creation:

“Not for too smart people who loves complex features and complex algorithms” a motivating quote found in the well known Machine Learning book by Bishop et. al. [8] provides the motivation to keep things simple if possible. This is exactly the back bone of the paper. The idea is to establish that if chosen in the proposed manner even simple statistical features can lead to good separability. Thus, eight most common statistical characteristics have been taken into account viz. Mean, Variance, Skewness, Kurtosis, Hyper-skewness, Hyper-flatness, Fano-Factor and Power Spectral Density. Each characteristic has been investigated for 3 mentioned frames leading to 24 Features. And the features are named as Mean-I, II, III; Variance-I, II, III, Skewness-I, II, III, Kurtosis-I, II, III, Hyper-skewness-I, II, III, Hyper-flatness-I, II, III, Fano-Factor-I, II, III and Power Spectral Density-I, II, III. Simple moments and spectral densities apparently look not too powerful or informative to be able to serve as good features for complicated non-trivial classification task like the one attempted here. The intuitive reasoning is that even if mean doesn’t serve as a good feature as a whole but if mean-I might be a good feature or say Skewness-II might be a good feature as will be investigated systematically in the subsequent sections with the help of adequate experimental results.

As mentioned before, a number of hit music of two both the decades in consideration are chosen according to the online public poles and rating sites. However, it is understandable that all the music especially those belonging to 90s are not available in a particular format. Also, as the music was gathered from different sources and music libraries, those differed in formats and the first pre-processing step was to convert all the songs to a common file format in order to process them. Thus, all the files were converted to .wave format in order to process them using MATLAB without much hustle. Once all the files were converted to the same format it was ready to process. As a starting point this paper has only considered Supervised Learning and thus all the processed file were labeled into either of the two categories Oldies or Contemporary. The above mentioned features are extracted from the audio signals generated from the songs. The description of each feature more particularly the mean and variance are provided in short for exploratory purpose in Table I to give an intuitive basic idea of the variability of the data.

Table.1. Description of the data points in the Feature Space

| Feature Description | Mean | | Variance | |
|---|---|---|---|---|
| | Class-I | Class-II | Class-I | Class-II |
| Mean –I | -0.00044438 | -0.0003612 | 9.21612719e-07 | 1.1215625588e-05 |
| Mean-II | -0.00072829 | -0.0004387 | 1.93245996e-06 | 1.1164467731e-05 |
| Mean-III | -0.00044737 | -0.0004999 | 8.98456077e-07 | 1.0865240822e-05 |
| Variance-I | 0.01204961 | 0.02626531 | 4.41411433e-05 | 0.00083378791069 |
| Variance-II | 0.02305717 | 0.06829872 | 7.03514107e-05 | 0.00137840159439 |
| Variance-III | 0.00905363 | 0.02675078 | 2.77991158e-05 | 0.00166861089596 |
| Skewness –I | 0.04611113 | -0.0273373 | 0.058800301377 | 0.06443525647532 |
| Skewness –II | 0.01273145 | 0.00198785 | 0.013608145790 | 0.00613314194182 |
| Skewness –III | -0.01777193 | 0.32556517 | 0.079012928726 | 5.57285477679514 |
| Kurtosis –I | 5.94044789 | 6.55330471 | 5.454891113407 | 33.3190301415651 |
| Kurtosis –II | 4.16133917 | 4.49096420 | 0.563421179411 | 7.82500140334606 |
| Kurtosis -III | 8.05177083 | 14.8732290 | 9.389433165898 | 2682.32876256885 |
| Hyperskewness-I | 0.01205069 | 0.02627642 | 4.414357877e-05 | 0.00083330569141 |
| Hyperskewness –II | 0.02305960 | 0.06830988 | 7.032234795e-05 | 0.00137753887067 |
| Hyperskewness –III | 0.00905469 | 0.02676162 | 2.780089400e-05 | 0.00166810867591 |
| Hyperflatness –I | -442.20255 | 1526.56723 | 74901927.8014086 | 95170936.0729312 |
| Hyperflatness –II | -1801.9041 | 1470.54026 | 24090913.5931019 | 252933565.716526 |
| Hyperflatness –III | -204.41265 | 1137.06389 | 2044493.05746511 | 69029260.7425727 |
| Fano Factor-I | 1.16145289 | -2.79265194 | 20.8693269740581 | 277.846136630811 |
| Fano Factor-II | 0.20466451 | 0.374239424 | 2.22660033200657 | 8.41749915674649 |
| Fano Factor-III | -0.3752638 | 154.2538851 | 60.3571735443054 | 1342204.15472221 |
| PSD –I | 81.0562016 | 178.9591291 | 7819.93310851926 | 255283.379816528 |
| PSD –II | 33.8941748 | 63.25612141 | 276.410742079874 | 23704.7632202317 |
| PSD -III | 146.340138 | 3726.054150 | 17660.9032223435 | 659559265.880186 |

## 4. Feature Selection:

“Not for too smart people who loves complex features and complex algorithms” a motivating quote found in the well known Machine Learning book by Bishop et.al. [8] provides the motivation to keep things simple if possible. Thus, though we have started with 24 Features, one might not want to use a 24 Dimensional Feature space and might want to reduce the dimensionality of the feature space and select most significant features and thus feature ranking has been introduced. A basic need to design a classifier with good generalization performance is that the number of training points must be large enough compared to the dimensionality of the feature space. It is a well known fact that under a limited number of training data, arbitrarily increasing the number of features leads to maximum possible value of error rate i.e. .5. In practice, increasing the number of features lead to initial improvement of performance but after a critical value further increase in dimensionality of feature space results in an increase in the error probability. This is known as peaking phenomenon [Raud 91, Duin 00]. This paper considers three different approaches in sequence to eliminate redundant features, rank them in the order of significance and reduce the dimensionality of the feature space taking into consideration only the most significant features to constitute the feature space used for classification purpose. The feature ranking and selection is discussed step by step as follows:

## 4.1. Feature Selection based on hypothesis testing:

The most natural first step would be to look at each of the generated feature in terms of their discriminatory power in case of the problem under consideration. Though looking at each feature independently is not optimal but this procedure helps rejecting 'bad' choices of features at the very beginning.

Hypothesis testing is a commonly used framework in Statistical Estimation and Detection theory to conclude on which hypothesis to select. For example if there are N observations $x_i, i = 1(1)N$ of the random variable x. $q = f(x_1, ...., x_N)$ be the selected function so that the probability density function of $q$ is easily parameterized by the unknown parameter $\theta$ i.e. $P_q(q;\theta)$. If $D$ be the interval of $q$ where it is more likely that $q$ lying under $H_0$ then if from the observed samples the obtain value of $q$ lies in $D$ then the null hypothesis is chosen else the alternate hypothesis is chosen. $D$ is known as the acceptance region and $\bar{D}$ is termed as the rejection region. In practice the probability of error when the null hypothesis is true is known as Significance Level which can be defined as: $P(q \in \bar{D} | H_0) \equiv \rho$. In practice, a particular value f the Significance Level is pre-selected. Assuming x to be a Gaussian random variable it can be shown that $q$ belongs to the so-called t-distribution with N-1 degrees of freedom. The interval values at different Significance Level and Degrees of Freedom are shown in Table.2. [Theodorodis.08].

| Degree of Freedom | $1-\rho$ | 0.9 | 0.95 | 0.975 | 0.99 | 0.995 |
|---|---|---|---|---|---|---|
| 10 | 1.81 | 2.23 | 2.63 | 2.63 | 3.17 | 3.58 |
| 11 | 1.79 | 2.20 | 2.59 | 2.59 | 3.10 | 3.50 |
| 12 | 1.78 | 2.18 | 2.56 | 2.56 | 3.05 | 3.43 |
| 13 | 1.77 | 2.16 | 2.53 | 2.53 | 3.01 | 3.37 |
| 14 | 1.76 | 2.15 | 2.51 | 2.51 | 2.98 | 3.33 |
| 15 | 1.75 | 2.13 | 2.49 | 2.49 | 2.95 | 3.29 |
| 16 | 1.75 | 2.12 | 2.47 | 2.47 | 2.92 | 3.25 |
| 17 | 1.74 | 2.11 | 2.46 | 2.46 | 2.90 | 3.22 |
| 18 | 1.73 | 2.10 | 2.44 | 2.44 | 2.88 | 3.20 |
| 19 | 1.73 | 2.09 | 2.43 | 2.43 | 2.86 | 3.17 |
| 20 | 1.72 | 2.09 | 2.42 | 2.42 | 2.84 | 3.15 |

While applying hypothesis testing in case of feature selection, the goal is to investigate whether the values a feature takes differ significantly over different classes. And, in the context of hypothesis testing, the problem is formulated as:

$H_1$: The values of the feature differ significantly over different classes (Alternate Hypothesis)
$H_0$: The value of the feature does not differ significantly. (Null Hypothesis)

This is approached by considering the differences in mean value of a corresponding feature in the two different classes and observed if these differences are significantly different to zero. Before processing all the features are made zero mean. Thus, the chosen test statistic is $q = \frac{(\bar{x} - \bar{y})}{s_z \sqrt{\frac{2}{N}}}$ where,

$s_z = \frac{1}{2N-2}(\hat{\sigma}_1^2 - \hat{\sigma}_2^2)$ Here, $\bar{x}$, $\bar{y}$, $\hat{\sigma}_1^2$, $\hat{\sigma}_2^2$ are respectively the mean for Class 1 and 2 and Variances for the two classes. However, these hypothesis based feature ranking or feature ranking based on Bhattacharya distance are seldom used in practical studies on Feature ranking. The main reason behind this is that these ranking schemes are not generalized and involves higher computational complexity and stronger assumptions. Thus a more widely used ranking scheme is used here as briefly described below with adequate justification of why FDR based ranking is preferred over these hypothesis test and Bhattacharya distance based feature ranking schemes.

### 4.2. Fisher's Discriminant Ratio (FDR):

In order to select most features effectively it is important to rank them to realize the contribution of each feature. The feature ranking scheme used here is based on Fisher Discriminant Ratio (FDR). FDR is a common measure to explore the discriminating power. Higher value of FDR is assigned to the features having higher difference in the mean value but smaller standard deviation implying compact yet distantly located clusters. Thus, features with higher FDR values are the most significant ones [3].

The main reason behind invoking yet another feature selection scheme in spite of having Bhattacharya Distance or hypothesis testing based feature separability measures is that all these previously mentioned methods are not easily computed, unless the Gaussian assumption is employed. Thus the requirement of a simpler criterion based on how the feature vector samples are scattered over the feature space. To measure this Within-Class Scatter matrix is defined as: $s_w = \sum_{i=1}^{M} P_i \Sigma_i$ where $\Sigma_i$ is the Covariance matrix for class $w_i$ where $\Sigma_i$ is the covariance matrix for class $w_i$ and $\Sigma_i = E[(x-\mu_i)(x-\mu_i)^T]$ and if $P_i$ is the a priori probability of class $w_i$ i.e. $P_i \approx n_i/N$ where $n_i$ is the number of samples in class $w_i$ out of total $N$ samples. Then the trace $\{s_w\}$ is obviously an average measure of the feature variance over all the classes. Two more measures are defined on the same philosophy- Between Class Scatter Matrix: $s_b = \sum_{i=1}^{M} P_i(\mu_i - \mu_0)(\mu_i - \mu_0)^T$ where, $\mu_o$ is the global mean vector. Mixture Scatter Matrix: $s_m = E[(x-\mu_0)(x-\mu_0)^T]$ the covariance matrix of the feature vector with respect to the global mean. Thus it follows straightway to combine these measures into a single measure as: $J_1 = \frac{trace\{s_m\}}{trace\{s_w\}}$ This leads to the straightforward conclusion that this measure takes higher values when within class samples are clustered together and between-class samples are well spread out. An alternate criterion can be used in the case of symmetric positive definite scatter matrices where the eigenvalues will be positive and thus trace is equal to sum of the eigenvalues leading to the justification of replacing the traces by determinants as the determinant is the product of the eigenvalues and hence large value of $J_1$ is same as the large value of $J_2$ defined as: $J_2 = \frac{|s_m|}{|s_w|}$ A variant of $J_2$ is often used in practice defined as: $J_3 = trace\{s_w^{-1} s_m\}$It can be shown easily [Theodorodis 08] that $J_2$ and $J_3$remain invariant under linear transformation thus make them advantageous to use. These criterions take a special form when considering two class problems. In this case it can be straightway seen that for equi-probable classes $|s_w| \propto (\sigma_1^2 + \sigma_2^2)$ and $|s_b| \propto (\mu_1 - \mu_2)^2$ Combining these two a new measure is defined as

$$FDR = \frac{(\mu_1 - \mu_2)^2}{(\sigma_1^2 + \sigma_2^2)}$$

Thus in the context of the problem in hand to get an idea about the features with best possible separability as well as independent of each other it is fairly natural to conduct a feature ranking using the FDR criterion. Thus, here on the basis of FDR all the 24 features in consideration are ranked in decreasing order of significance i.e. in the decreasing order of FDR value as it can be fairly understood that higher the value of FDR the more significant the feature is. This is because higher value of FDR means the inter-class separation is larger however the between class separation is smaller i.e. the samples of the feature same class are well clustered together and the feature samples of different classes are well separated. The complete ranking of the 24 features based on FDR scores are given in Table.3. Also the relative significance of one feature to the other can be readily identified from their relative values as visualized from the Fig.4.in decreasing order of importance. It can be seen easily that there exist six groups of features. Thus features belonging to a common group have similar importance and thus the features should be considered or eliminated in group as that makes more sense.

Table.3. Feature Ranking using FDR score in the order of decreasing importance

| Feature Ranking | FDR | Feature Description |
|---|---|---|
| $F_1$ | 9.2880 | Power Spectral Density of Frame 2 |
| $F_2$ | 9.2834 | Variance of Frame 2 |
| $F_3$ | 3.7508 | Power Spectral Density of Frame 1 |
| $F_4$ | 3.7472 | Variance of Frame 1 |
| $F_5$ | 3.3582 | Power Spectral Density of Frame 3 |
| $F_6$ | 3.3559 | Variance of Frame 3 |
| $F_7$ | 1.7868 | Skewness of Frame 1 |
| $F_8$ | 1.6341 | Hyper-Skewness of Frame 1 |
| $F_9$ | 1.5356 | Fano- factor of Frame 2 |
| $F_{10}$ | 1.4907 | Skewness of Frame 2 |
| $F_{11}$ | 1.4809 | Hyper Flatness of Frame 2 |
| $F_{12}$ | 1.2428 | Hyper Flatness of Frame 1 |
| $F_{13}$ | 1.1791 | Fano- factor of Frame 3 |
| $F_{14}$ | 1.1279 | Fano- factor of Frame 1 |
| $F_{15}$ | 1.0886 | Hyper Flatness of Frame 3 |
| $F_{16}$ | 1.0424 | Hyper Skewness of Frame 3 |
| $F_{17}$ | 1.0269 | Kurtosis of Frame 3 |
| $F_{18}$ | 0.8889 | Skewness of Frame 3 |
| $F_{19}$ | 0.7687 | Mean of Frame 3 |
| $F_{20}$ | 0.6248 | Kurtosis of Frame 2 |
| $F_{21}$ | 0.5972 | Mean of Frame 2 |
| $F_{22}$ | 0.4059 | Mean of Frame 1 |
| $F_{23}$ | 0.1864 | Kurtosis of Frame 1 |
| $F_{24}$ | 0.1197 | Hyper Skewness of Frame 2 |

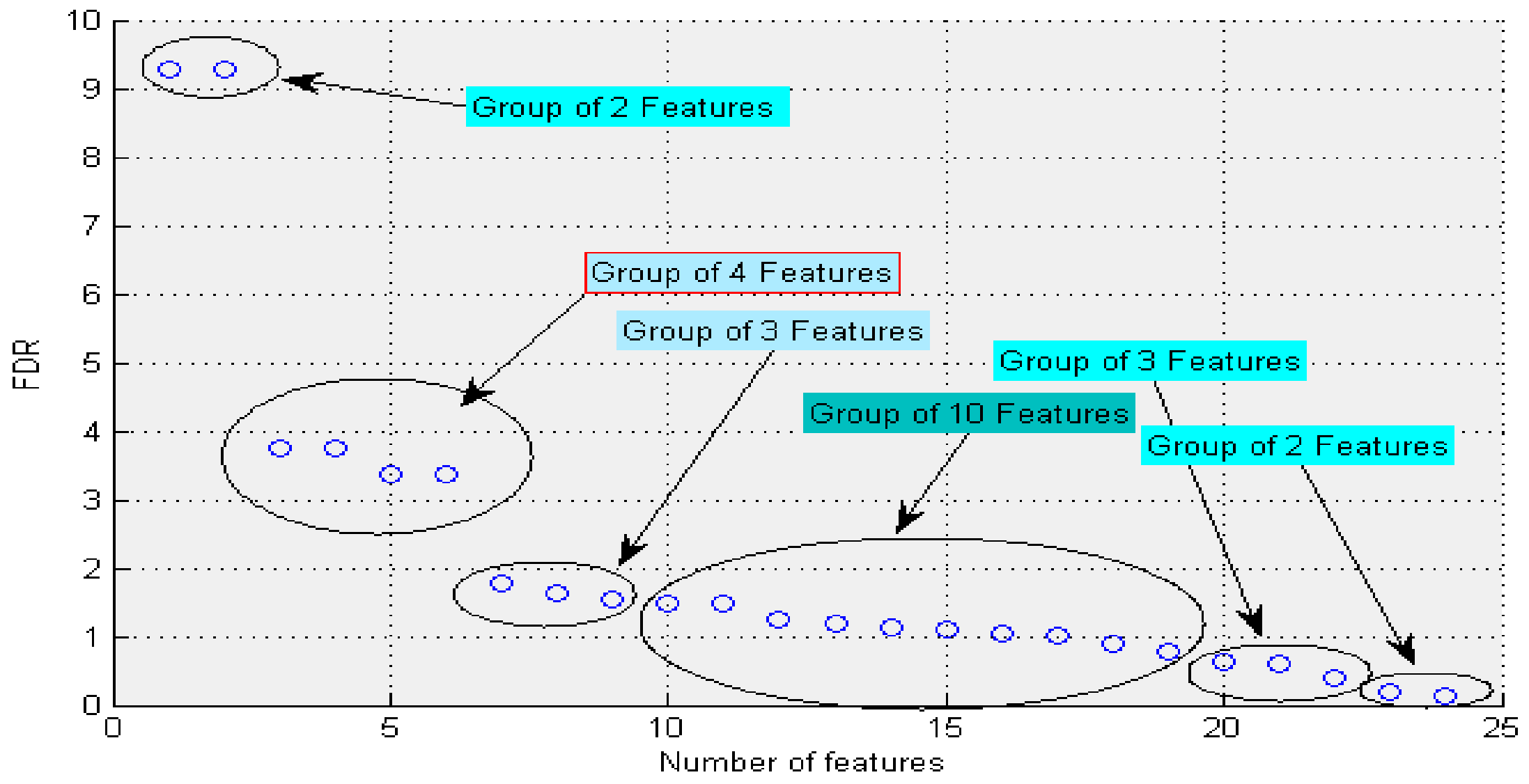

Fig.4. Feature Ranking using FDR

Thus to look into the separability power of these features we look into the 2D scatter plots between F1-F2, F2-F3 and F3-F1 also we look at the 3D scatter plots among the top 4 features as F1-F2-F3, F1-F2-F4,F1-F3-F4,F2-F3-F4 as shown in Fig.5- Fig. 12.

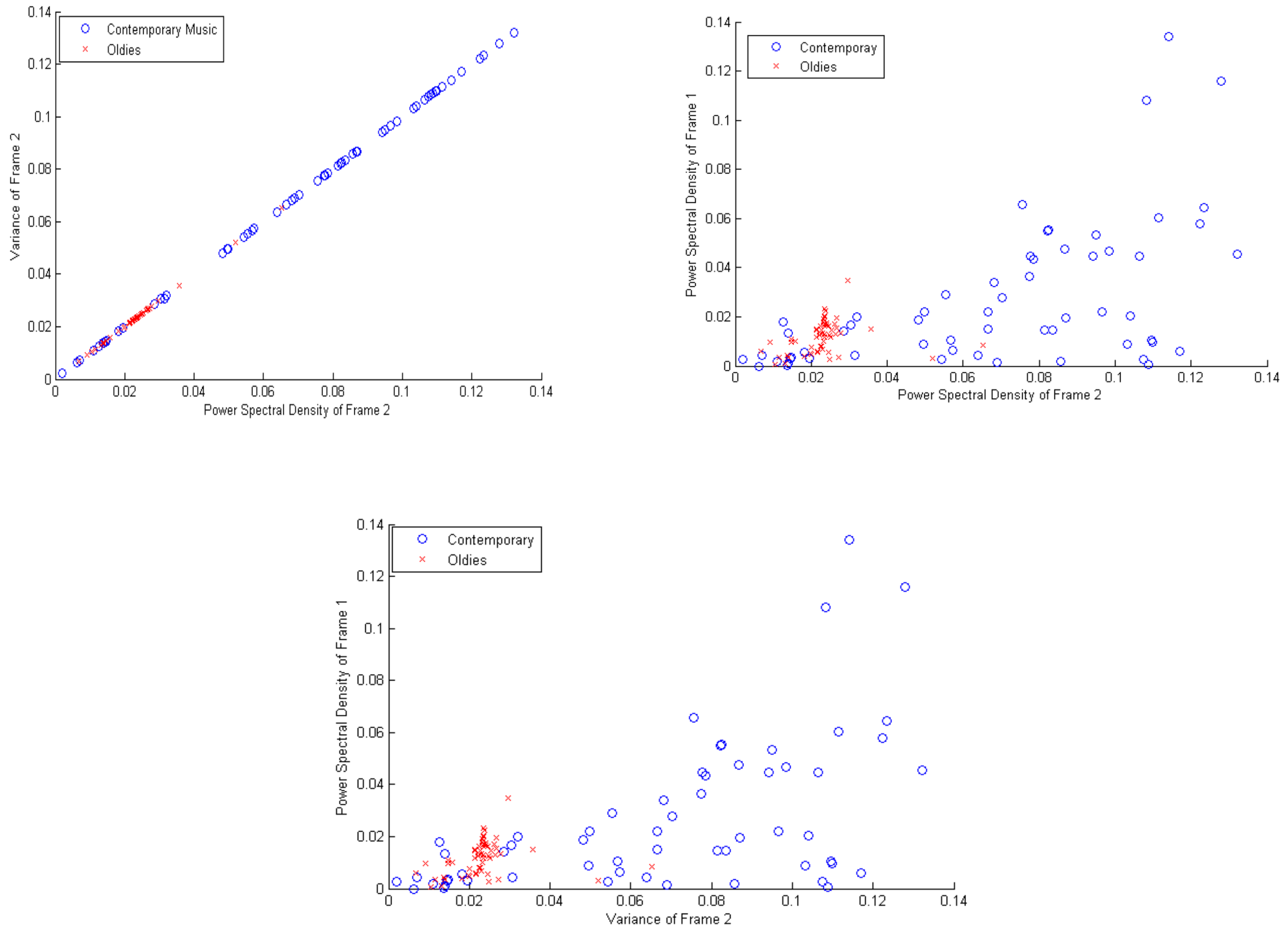

Fig.5, 6, 7.   2D Scatter Diagram of top 3 Features

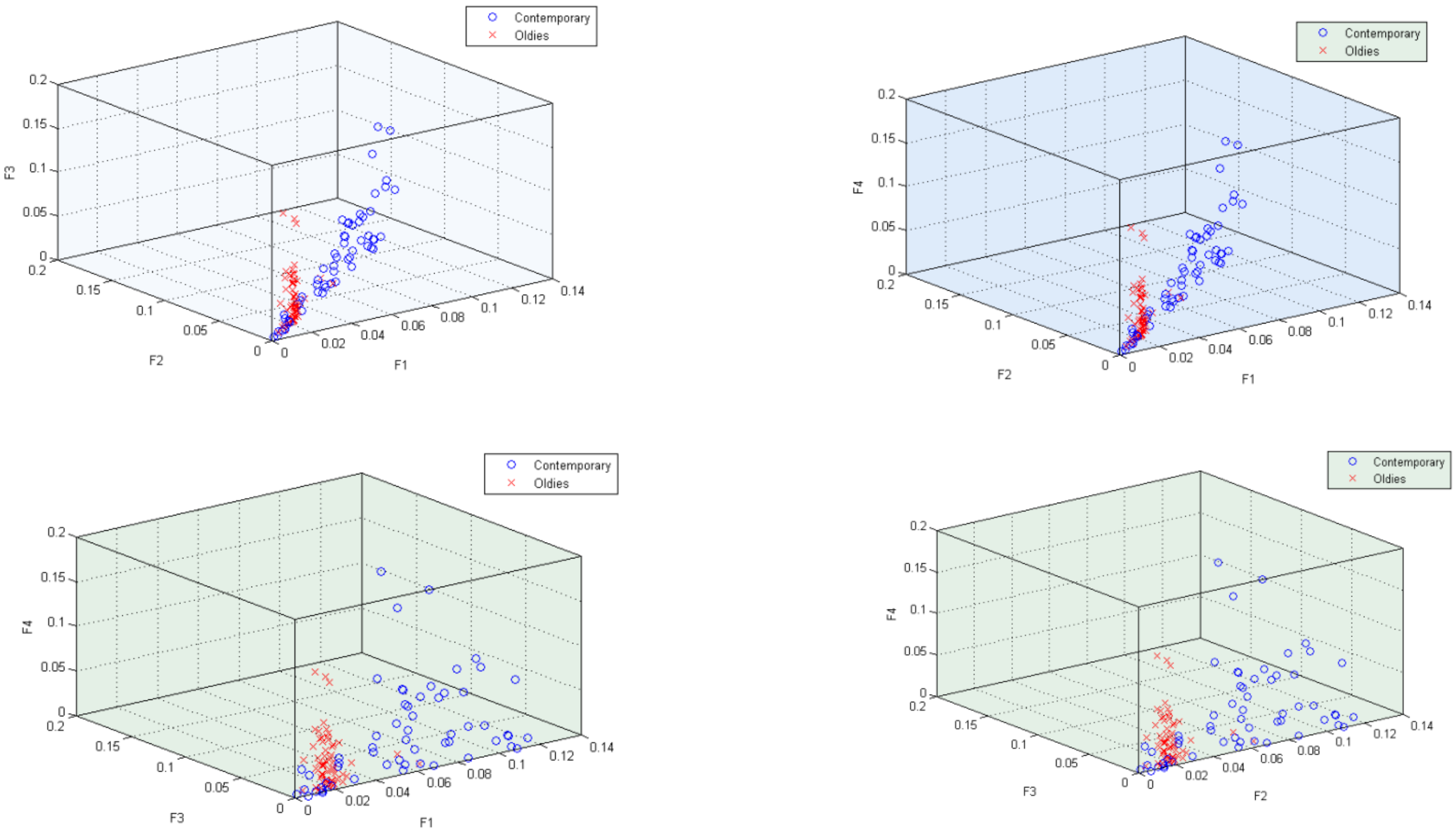

Fig. 8, 9, 10, 11. 3D scatter diagram for top four features for different class

**4.3. Singular Value Decomposition/ Principle Component Analysis and Feature Generation:**

Principal component analysis (PCA) is a statistical procedure that uses orthogonal transformation to convert a set of observations of possibly correlated variables into a set of values of linearly uncorrelated variables called principal components. The number of principal components is less than or equal to the number of original variables. This transformation is defined in such a way that the first principal component has the largest possible variance (that is, accounts for as much of the variability in the data as possible), and each succeeding component in turn has the highest variance possible under the constraint that it is orthogonal to (i.e., uncorrelated with) the preceding components. Principal components are guaranteed to be independent if the data set is jointly normally distributed. PCA is sensitive to the relative scaling of the original variables.

Here in order to select the best possible uncorrelated features we used PCA based feature generation technique. As we ran PCA over the feature space and obtained the combined feature space in terms of the Principal Components. Note that, here newly formed Principal Components (PC) are treated as features as all our chosen features have been embedded into this PCs in an uncorrelated manner. Table4. Presents the variances explained by each principal Component and Fig 12 presents the corresponding Scree Plot.

Table4. Variance Explained in the direction of PCs

| Variance Explained (in %) | No. of Principle Components |
|---|---|
| 19.0991166374520 | PC 1 |
| 35.8146375252128 | PC 1-2 |
| 49.1015387122730 | PC 1-3 |
| 61.2559301916625 | PC 1-4 |
| 70.5946491487700 | PC 1-5 |
| 78.0352244355646 | PC 1-6 |
| 83.1832352675688 | PC 1-7 |
| 87.6578217512420 | PC 1-8 |
| 91.9040716200055 | PC 1-9 |
| 95.0081327487821 | PC 1-10 |
| 97.3108230132306 | PC 1-11 |
| 98.4159143262898 | PC 1-12 |
| 99.0393002618932 | PC 1-13 |
| 99.5955565572968 | PC 1-14 |
| 99.7361183548248 | PC 1-15 |
| 99.8215038586424 | PC 1-16 |
| 99.8877898964263 | PC 1-17 |
| 99.9422300035963 | PC 1-18 |
| 99.9805629558746 | PC 1-19 |
| 99.9995059051504 | PC 1-20 |
| 99.9999986604658 | PC 1-21 |
| 99.9999999938943 | PC 1-22 |
| 99.9999999979122 | PC 1-23 |
| 100 | All 24 PC |

Fig12. Scree Plot indicating relative importance of the Components

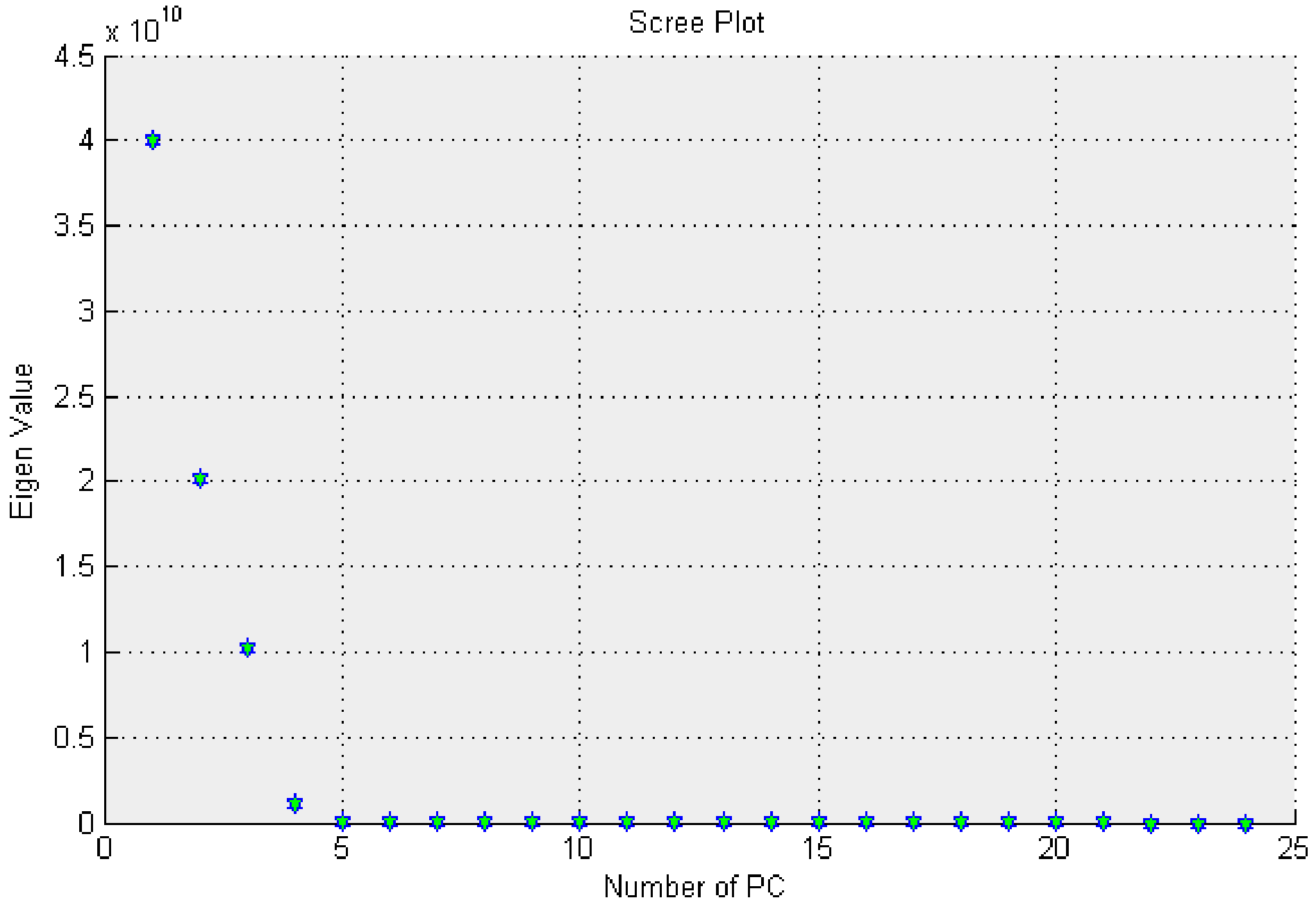

### 5. Classification Performance:

In order to establish any claims regarding the performance of the proposed algorithm of dividing the song into three mentioned frames to extract features, the only way is to prove that is to show in terms of classification performance. And further in order to establish the claim that it performs better than classification based on usual feature extraction scheme using the same set of features but extracted in the usual manner i.e. from the entire signal as a whole a thorough step by step comparison is presented.

The classification is performed in a Sequential Forward Selection (SFS) manner. Also, to ensure generalized result and minimize the effect of over-fitting the classification tasks are carried out in a monte- carlo manner 500 iterations for each classifier.

The results are presented in three sub-sections. First, we provide classification results for the proposed algorithm and a comparative study with the classification performance for the case where the same basic features are chosen but in the usual manner i.e. from the entire audio signal as a whole. Thus, it has been shown that the proposed algorithm increases accuracy almost 10-15%. Here, the feature selection is done in accordance with the FDR score and employing SFS over that. Secondly, the similar comparative study has been done on the basis of the linearly independent hybrid features generated using SVD based dimensionality reduction for both the proposed and existing method in a similar SFS manner.

As mentioned before all the classification results have been generated using 500 fold cross validation in a monte-carlo manner thus ensuring generalization and eliminating the effect of over fitting.

Table5. Classification performance based on FDR based Feature Selection

| Classifier | Variants | Feature Set | Misclassification Rate |
|---|---|---|---|
| Discriminant Analysis | Linear | Top 2 | 0.1628 |
| | | Top 6 | 0.1575 |
| | | Top 9 | 0.1750 |
| | | Top 19 | 0.1955 |
| | | Top 22 | 0.1970 |
| | | All 24 | 0.2105 |
| | Diaglinear | Top 2 | 0.1695 |
| | | Top 6 | 0.1858 |
| | | Top 9 | 0.1675 |
| | | Top 19 | 0.1776 |
| | | Top 22 | 0.1788 |
| | | All 24 | 0.1839 |
| | QDA | Top 2 | 0.1382 |
| | | Top 6 | 0.0972 |
| | | Top 9 | 0.1015 |
| | | Top 19 | 0.1990 |
| | | Top 22 | 0.2204 |
| | | All 24 | 0.2204 |
| | Diagquadratic | Top 2 | 0.1461 |
| | | Top 6 | 0.1561 |
| | | Top 9 | 0.1694 |
| | | Top 19 | 0.1843 |
| | | Top 22 | 0.1782 |
| | | All 24 | 0.1853 |
| | Mahalanobis | Top 2 | 0.1078 |
| | | Top 6 | 0.2678 |
| | | Top 9 | 0.3506 |
| | | Top 19 | 0.3871 |
| | | Top 22 | 0.3956 |
| | | All 24 | 0.3956 |
| Support Vector Machine | Linear Kernel | Top 2 | 0.1654 |
| | | Top 6 | 0.1656 |
| | | Top 9 | 0.1790 |
| | | Top 19 | 0.1912 |
| | | Top 22 | 0.1772 |
| | | All 24 | 0.1872 |
| | Quadratic Kernel | Top 2 | 0.1673 |
| | | Top 6 | 0.1735 |
| | | Top 9 | 0.2067 |
| | | Top 19 | 0.2327 |
| | | Top 22 | 0.2134 |
| | | All 24 | 0.2184 |
| | Rbf Kernel | Top 2 | 0.1418 |
| | | Top 6 | 0.1347 |
| | | Top 9 | 0.1548 |
| | | Top 19 | 0.1353 |
| | | Top 22 | 0.1290 |
| | | All 24 | 0.1375 |

Table6. Classification performance based on PCA based Features

| Classifier | Variants | Feature Set | Misclassification Rate |
|---|---|---|---|
| Discriminant Analysis | Linear | PC 1-5 | 0.1605 |
| | | PC 1-7 | 0.1626 |
| | | PC 1-9 | 0.1646 |
| | | PC 1-11 | 0.1614 |
| | | PC 1-13 | 0.1818 |
| | | All 24 PC | 0.1874 |
| | Diaglinear | PC 1-5 | 0.1925 |
| | | PC 1-7 | 0.1984 |
| | | PC 1-9 | 0.1999 |
| | | PC 1-11 | 0.1706 |
| | | PC 1-13 | 0.1851 |
| | | All 24 PC | 0.1934 |
| | QDA | PC 1-5 | 0.1626 |
| | | PC 1-7 | 0.1656 |
| | | PC 1-9 | 0.1611 |
| | | PC 1-11 | 0.1656 |
| | | PC 1-13 | 0.1910 |
| | | All 24 PC | 0.2006 |
| | Diagquadratic | PC 1-5 | 0.2087 |
| | | PC 1-7 | 0.2173 |
| | | PC 1-9 | 0.2311 |
| | | PC 1-11 | 0.1923 |
| | | PC 1-13 | 0.1959 |
| | | All 24 PC | 0.2156 |
| | Mahalanobis | PC 1-5 | 0.4167 |
| | | PC 1-7 | 0.4549 |
| | | PC 1-9 | 0.4154 |
| | | PC 1-11 | 0.3815 |
| | | PC 1-13 | 0.3708 |
| | | All 24 PC | 0.2993 |
| Support Vector Machine | Linear Kernel | PC 1-5 | 0.1479 |
| | | PC 1-7 | 0.1492 |
| | | PC 1-9 | 0.1616 |
| | | PC 1-11 | 0.1620 |
| | | PC 1-13 | 0.1818 |
| | | All 24 PC | 0.1831 |
| | Quadratic Kernel | PC 1-5 | 0.1509 |
| | | PC 1-7 | 0.1770 |
| | | PC 1-9 | 0.1926 |
| | | PC 1-11 | 0.2071 |
| | | PC 1-13 | 0.2289 |
| | | All 24 PC | 0.2725 |
| | Rbf Kernel | PC 1-5 | 0.1198 |
| | | PC 1-7 | 0.1289 |
| | | PC 1-9 | 0.1460 |
| | | PC 1-11 | 0.1347 |
| | | PC 1-13 | 0.1442 |
| | | All 24 PC | 0.1475 |

As can be seen from the Classification performance that both the methods did fine and FDR based techniques slightly did better than PCA based technique. As can be seen that FDR based Classification performs best with top 6 features thus we might conclude that after that it is going to over-fitting regime. Also, these Classifiers were run in a sequential Monte- Carlo manner thus taking 500 fold cross validation into account. Also, Classification results are best and almost around 88% accuracy is obtained with top 5-7 Principle Components. Whereas, FDA based method reported a best accuracy of 91% with top 6 features. Among the Classifiers LDA, QDA and SVM performance were pretty similar however, QDA and SVM (especially Radial Basis Function as Kernel) performed better than others. Table no. 5 and 6 presents the Classification Performances in a compact manner.

## 6. Conclusion:

The approach to classify musical contents in this paper is philosophically different from classical approaches. The Classical approach looks at the entire signal but here we have proposed a possible new scheme to analyze audio content, splitting the audio signal in three particular frames which is far simpler than sliding frame models as in case of sliding window scheme one will get continuous series of features which needs to be segmented further for classification. Also potentially more powerful as compared to extracting features from the entire signal at a go as this is like zooming in different portions of the audio signal and extracting each most powerful features from each segment.

Also, the dataset created and analyzed here is different than the existing ones and this study might be significant as it points to a possible statistical change in the Musical Taste of Music Listeners over a decade. Hit Music of two subsequent decades seems so different which looks like an interesting observation and can be extended further to regression problems and it might be interesting to predict the future trend of Hit Music statistically. This might be an interesting thing to look into and we left it as a future work.

Also, here as we approached the Classification task from two different approaches relying on FDR and PCA based Feature Selection strategy. Both of the approaches performs fairly well however, FDR based approach seems to perform slightly better. Typical choice of top 6-7 Features based on FDR gives about 91% accuracy over different classifiers whereas the best performance reported in PCA based feature selection is around 88% taking top 7 Principle Components as features. This might be a possible finding that under the proposed framework FDR based approach performs better than PCA based approach which leads to another possible comment about the Trapezoidal method, that it might leads to finding good and mostly significant features as compared to classical approaches.

**Acknowledgement:**

This project is an outcome of UCI CS277 Course. I would like to thank Professor Padhraich Smyth for offering CS277 and making it fun and flexible- an open environment to learn and more importantly to think in a new way.